\documentstyle[aps,epsfig,multicol]{revtex}
\draft
\input epsf
\begin{document}
\title{\bf Nonlinear dynamics and surface diffusion of diatomic molecules}
\author{C. Fusco${}^{*[{\rm a}]}$ and A. Fasolino${}^{[{\rm a}],[{\rm b}]}$\\
${}^{[{\rm a}]}$ Theoretical Physics, Institute for Molecules and Materials,\\
Radboud University Nijmegen,\\ 
Toernooiveld 1, 6525 ED Nijmegen (The Netherlands)\\
${}^{[{\rm b}]}$ HIMS/WZI, Faculty of Science,\\
University of Amsterdam,\\
Nieuwe Achtergracht 166,
1018 WV Amsterdam (The Netherlands)\\
}

\date{\today}
\maketitle

\begin{abstract}
The motion of molecules on solid surfaces is of interest for 
technological applications, but it is also a theoretical challenge.
We study the deterministic and thermal diffusive dynamics of a dimer
moving on a periodic substrate. The deterministic motion of the dimer 
displays strongly nonlinear features and chaotic behavior.
The dimer thermal diffusive dynamics 
deviates from simple Arrhenius behavior, due to the coupling between 
vibrational and translational degrees of freedom.
In the low-temperature limit the dimer diffusion can become orders
of magnitude larger than that of a single atom, as also found experimentally.
The relation between chaotic deterministic 
dynamics and stochastic thermal diffusion is discussed.
\end{abstract}
\vspace*{0.5cm}
keywords: Surface, Molecules, Molecular Dynamics

\begin{multicols}{2}

\section{Introduction}
\label{sec.intro}

The surface diffusion of single adatoms has been intensively studied 
over the last decades,~\cite{b.risken,b.kellogg,b.ala}
due to its importance in thin film and crystal growth.~\cite{b.venables}
Once individual atoms are adsorbed on a surface they can meet, thus forming 
larger clusters.
However, the diffusion of even the simplest cluster, a dimer, 
on a surface is by far not yet understood.~\cite{b.wang,b.braun,b.montalenti1,b.boisvert,b.kovalev,b.romero,b.goncalves,b.fusco1,b.fusco2}
The diffusion dynamics can be strongly affected by the coupling of the 
intramolecular motion to the translational motion of the centre of mass (CM)
of the cluster.~\cite{b.fusco1,b.fusco2,b.deltour,b.krylov}
Herein, we present a simple, one-dimensional model for studying the Hamiltonian
and diffusive dynamics of a dimer, which is relevant to systems where 
quasi-one-dimensional motion takes place.~\cite{b.kurpick} 
The deterministic dynamics of this model is characterized by a complex 
behavior, dominated by non-linear effects, parametric resonances and chaotic 
features. 
At variance with the case of a single atom, at $T\ne 0$ the role of the 
internal degrees of freedom of the dimer is responsible for deviations from 
activated behavior of the diffusion coefficient.
In Section~\ref{sec.model} we briefly outline our model. 
In Sections~\ref{sec.deterministic} and~\ref{sec.thermal} we discuss 
the nonlinear deterministic and thermal dynamics, respectively, and compare 
the two situations in Section~\ref{sec.relation}. 
Concluding remarks are given in Section~\ref{sec.conclusion}.

\section{Model}
\label{sec.model}
We consider the deterministic and thermal dynamics of a dimer moving on a 
periodic one-dimensional substrate. The particle-substrate interaction is a 
sinusoidal function of amplitude $2U_0$ and period $a$, and the 
interparticle interaction is given by a harmonic potential with spring 
constant $K$ and equilibrium length $l$.
We use Langevin dynamics to deal with finite temperature $T$. 
The equations of motion for the two atoms of mass $m$ and of coordinates
$x_1$ and $x_2$ composing the dimer are given by Equation~(\ref{e.dim}):
\begin{equation}
\label{e.dim}
\left\{ \begin{array}{ccc}
m\ddot{x}_1 +m\eta\dot{x}_1 & = & K(x_2-x_1-l)-\frac{2\pi U_0}{a}
\sin\left(\frac{2 \pi x_1}{a}\right)+f_1\\
m\ddot{x}_2 +m\eta\dot{x}_2 & = & K(x_1-x_2+l)-\frac{2\pi U_0}{a}
\sin\left(\frac{2\pi x_2}{a}\right)+f_2
\end{array} \right.
\end{equation}
where the effect of finite temperature $T$ is taken into account 
by the stochastically fluctuating forces $f_i$, satisfying the conditions
$<f_i(t)>=0$ and $<f_i(t)f_j(0)>=2m\eta k_BT\delta_{ij}\delta(t)$, and by
the damping term $m\eta\dot{x}_{i}$.
In the following, we will use representative values of the parameters, 
$a=0.25$ nm, $U_0=0.2$ eV, $m=5\times 10^{-26}$ kg, $\eta=0.7$ ps$^{-1}$. 
The values of $K$ and $l$ will be given in the caption of each figure.
We performed molecular dynamics (MD) simulations, integrating the equations of 
motion using a velocity-Verlet algorithm, with a
time step $\Delta=10^{-16}$ s and averaging the trajectories over several 
thousands of realizations in the case of thermal diffusion, in order to 
reduce the statistical noise.

\section{Hamiltonian Dynamics}
\label{sec.deterministic}

\begin{figure}
\centering\epsfig{file=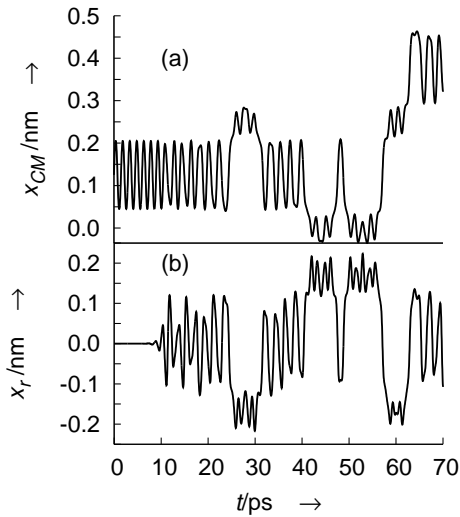,scale=0.8}
\caption{Dynamics given by Equation~(\ref{e.CM}) for 
$K=0.2$ Nm$^{-1}$ and $v_0=380$ ms$^{-1}$, $f_i=0$ and $\eta=0$. 
(a) The CM motion is plotted; (b) the deviations from equilibrium of the
internal coordinate are plotted.}
\label{f.dimerdetU_00.6K0.1eta0l1}
\end{figure}

First we consider the Hamiltonian dynamics 
($f_i=0$ and $\eta=0$ in Equation~(\ref{e.dim})).
It is convenient to rewrite Equation~(\ref{e.dim}) in terms of 
the CM coordinate $x_{CM}=(x_1+x_2)/2$ and of the deviations from equilibrium 
of the internal coordinate $x_r=x_2-x_1-l$, obtaining Equation~(\ref{e.CM}):
\begin{equation}
\label{e.CM}
\left\{ \begin{array}{lll}
m\ddot{x}_{CM} & = & -\frac{2\pi U_0}{a}
\sin\left(\frac{2\pi x_{CM}}{a}\right)
\cos\left[\frac{\pi}{a} \left(x_r+l\right)\right] \\
m\ddot{x}_r & = & -2Kx_r-\frac{2\pi U_0}{a}
\cos\left(\frac{2\pi x_{CM}}{a}\right)
\sin\left[\frac{\pi}{a} \left(x_r+l\right)\right]
\end{array} \right.	
\end{equation}
We have considered the case of a commensurate dimer 
($l=a$) starting at equilibrium with a given initial kinetic energy
$E_{kin}^0$, a case which allows some analytical results for the 
initial phase of the motion, showing the role of internal vibrations on the 
dynamics, to be obtained.~\cite{b.fusco1}
In fact, for a rigid dimer with $\dot{x}_1(0)=\dot{x}_2(0)=v_0$, 
the minimum  kinetic energy for the CM to overcome 
the potential barrier is $E_{kin}^0=mv_0^2=4U_0$.
Hence, for $v_0<\sqrt{4U_0/m}$, the motion of the CM is oscillatory,
while a drift regime is attained for $v_0>\sqrt{4U_0/m}$.
Conversely, when the dimer is allowed to vibrate, the coupling between the CM
and the internal motion makes it possible for the CM of the dimer to overcome
the potential barrier $4U_0$ for values of $v_0$ below the threshold
$\sqrt{4U_0/m}$. 
In fact, if the internal motion is excited, 
it can happen that one particle remains in the minimum and the other reaches 
the nearest maximum. 
From the energy balance, Equation~(\ref{e.energybal}) follows: 
\begin{equation}
\label{e.energybal}
E_{kin}^0=2U_0+\frac{1}{2}K(a/2)^2
\end{equation}
If $K$ is sufficiently small, the right-hand side of 
Equation~(\ref{e.energybal}) can be smaller 
than $4U_0$. This is the situation shown in 
Figure~\ref{f.dimerdetU_00.6K0.1eta0l1}, where the CM motion is rather 
irregular, behaving in a chaotic fashion.
\noindent
This chaotic regime occurs for weakly bound dimers in a velocity window 
around the threshold $\sqrt{4U_0}$, and can be characterized by Lyapunov 
exponents and power spectra.~\cite{b.fusco1}
We have also shown that, 
for larger values of $K$, the drift motion of the CM may excite the internal 
vibrations by a parametric resonance in a 
velocity window around twice the natural stretching frequency of the dimer 
$\omega_0=\sqrt{2K/m}$. 

\section{Thermal Diffusion}
\label{sec.thermal}

\begin{figure}
\centering\epsfig{file=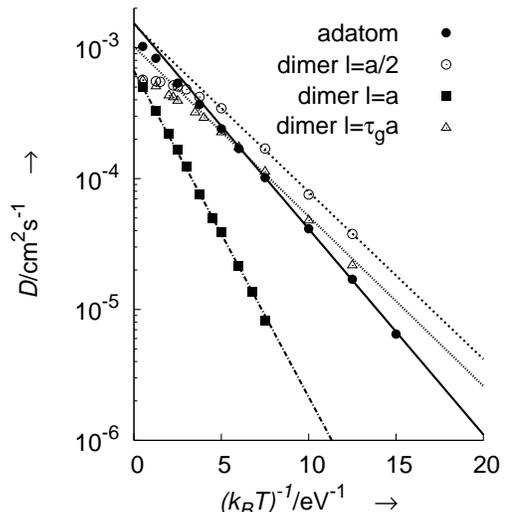,scale=0.8}
\caption{Diffusion coefficient $D$ as a function of 
$1/(k_BT)$ for the adatom and the dimers with different values of $l$ 
and $K=2$ Nm$^{-1}$. The points are the result of the simulations and the  
lines represent fits to the data in the low-temperature regime.}
\label{f.diffusmon-l0.5-l1-inc}
\end{figure}

The thermal diffusive behavior of the dimer is characterized by 
computing the diffusion coefficient $D$ from the mean square displacement 
$<x_{CM}^2(t)>$, as in Equation~(\ref{e.diffcoeff}):
\begin{equation}
\label{e.diffcoeff}
D=\lim_{t\rightarrow\infty}\frac{<x_{CM}^2(t)>}{2t}.
\end{equation}
For temperatures that are small compared to the energy barrier, Kramer's 
theory~\cite{b.hanggi} predicts
an activated Arrhenius behavior of diffusion, given by 
Equation~(\ref{e.arrhenius}):
\begin{equation}
\label{e.arrhenius}
D=D_0\exp(-E_a/k_BT),
\end{equation} 
where the activation energy $E_a$ and the prefactor $D_0$ do not depend on 
$T$. In the limit of vanishing energy barrier, the adatom diffusion 
coefficient obeys the Einstein relation and is given by $D=k_BT/(m\eta)$, 
twice the value of the dimer $D=k_BT/(2m\eta)$. The diffusion coefficient of 
a single adatom shown in Figure~\ref{f.diffusmon-l0.5-l1-inc} has an
activation energy corresponding to the energy barrier $2U_0=0.4$~eV, 
except at very high temperatures owing to finite barrier 
effects.~\cite{b.montalenti2}
\noindent
For a rigid dimer, $E_a=4U_0$ for $l=a$ down to $E_a=0$ for
$l=a/2$.  If the dimer is not rigid, the activation energy is a
non trivial function of the equilibrium length and elastic constant. 
In Figure~\ref{f.diffusmon-l0.5-l1-inc}, we show the diffusion 
coefficient for non rigid dimers with different values of $l$. 
We find $E_a=0.6$~eV for $l=a$, $E_a=0.3$ eV for $l=a/2$ and 
$E_a=0.34$~eV for the incommensurate case $l=\tau_g a$, where
$\tau_g=(1+\sqrt{5})/2$ is the golden mean. The finite-barrier corrections to 
the Arrhenius behavior at high temperatures ($k_BT>2U_0$) are rather 
pronounced for $l\ne a$. 
Deviations from the Arrhenius law due to incommensurability are also found 
for larger clusters.~\cite{b.hamilton}

\begin{figure}
\centering\epsfig{file=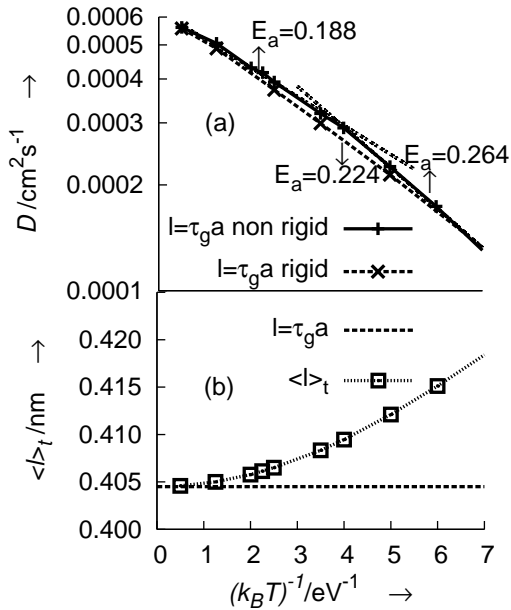,scale=1}
\caption{(a) Diffusion coefficient as a function of 
$1/(k_BT)$ for the incommensurate dimer with 
$l=\tau_ga$ and $K=2$ Nm$^{-1}$. The intramolecular length was 
kept fixed or not fixed, as indicated by the labels. 
Fitted activation energies [eV] are also reported. (b) 
Dynamical equilibrium length $<l>_t$ as a function of $1/(k_BT)$.} 
\label{f.diffusinc}
\end{figure}

It has been suggested that non-Arrhenius behavior can result also from 
dynamical effects related to the internal motion of the 
dimer.~\cite{b.fusco2,b.krylov} The role of the internal vibrations on the 
diffusive behavior is illustrated in Figure~\ref{f.diffusinc}(a), where we 
compare the diffusion coefficient of the rigid and non rigid incommensurate 
dimer ($l=\tau_ga$).

It is clear that we can define a unique value of $E_a$ for the rigid dimer, 
whereas the activation energy is, in general, temperature dependent 
when the dimer is allowed to vibrate. 
The temperature dependence of $E_a$ is linked to a temperature dependent 
misfit $<l>_t$, induced by the dynamics, as shown in 
Figure~\ref{f.diffusinc}(b).

Moreover, the dimer can even diffuse faster than the adatom, at least for 
$l\neq a$ and low temperatures. 
Enhanced diffusivity of dimers and small clusters is also found by theoretical
studies of $1D$ diffusion in molecular sieves~\cite{b.sholl1} and in zeolite
crystals.~\cite{b.sholl2}
Mitsui et al.~\cite{b.mitsui} measured water
diffusion on Pd(111) at low temperature ($T\simeq 40$ K), finding the 
mobility of dimers and larger clusters to
be $3-4$ orders of magnitude larger than that of adatoms. 
This experimental result is compatible with our findings. In fact, 
extrapolation to low temperature of our results of 
Figure~\ref{f.diffusmon-l0.5-l1-inc} shows that the dimer diffusion
for $l\ne a$ can be orders of magnitude higher than for the adatom.
pin

\section{Relation between Deterministic and Thermal Diffusion}
\label{sec.relation}

\begin{figure}
\centering\epsfig{file=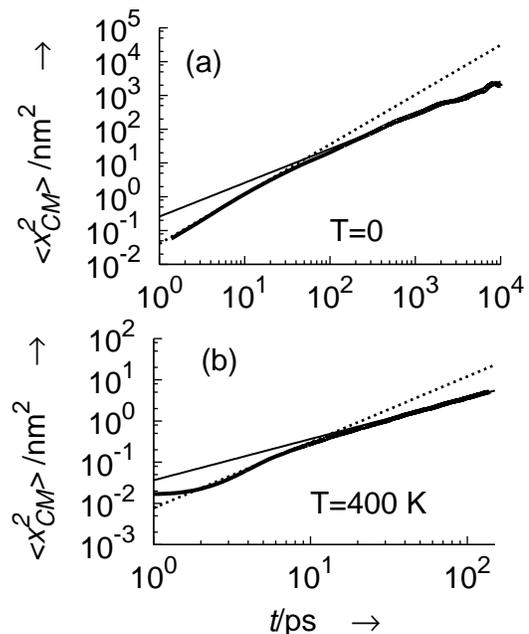,scale=1}
\caption{Mean-square displacement for (a) the deterministic and (b)
the thermal motion for $l=a$ (thick solid lines). 
(a) $K=0.1$ Nm$^{-1}$, $v_0=475$ ms$^{-1}$, $\eta=0$. (b) $K=0.2$ Nm$^{-1}$. 
The thin solid lines are linear fits for large $t$;
the dotted lines are power-law fits with exponent $\sim 1.5$ for small
$t$. The data plotted in (b) were obtained by averaging the trajectory 
over $3000$ realizations.}
\label{f.x2}
\end{figure}

We find that the chaotic dynamics discussed in 
Section~\ref{sec.deterministic} can give rise to a diffusive behavior, even in 
the absence of thermal fluctuations. 
The role of the heat bath is played by the exchange between 
translational and internal motion which, owing to the nonlinearity of the 
system, can occur in a random manner. Figure~\ref{f.x2} shows a comparison 
between the deterministic and the thermal mean square displacements. 
\noindent
In the deterministic case $<x_{CM}^2>$ represents a time averaging taken by 
displacing the time origin,~\cite{b.allen} whereas at $T\ne 0$ it is an average
over realizations.
The long-time behavior is linear (diffusive) even for the Hamiltonian 
dynamics in the chaotic regime. In both cases we also
find a transient superdiffusive regime ($<x^2_{CM}>\propto t^{\alpha}$, with 
$\alpha\simeq 1.5$). Anomalous diffusion with $\alpha\simeq 7/5$ 
has also been observed in a model of adatom surface diffusion in 
two dimensions.~\cite{b.guantes} Anomalous diffusion might be related to long 
jumps of the dimer, that is to trajectories that move over multiple surface 
minima,~\cite{b.guantes,b.sholl3} which do occur during our simulations.
Furthermore, it has been claimed that a 
deterministic diffusive behavior leads to a non-Arrhenius dependence of the 
thermal diffusion coefficient,~\cite{b.kopelevich} which is compatible with 
our results.
\section{Conclusions}
\label{sec.conclusion}

We have presented a one-dimensional model to describe the diffusive dynamics
of dimers on periodic surfaces. 
We have shown that the coupling between translational and vibrational degrees
of freedom can lead to Hamiltonian chaotic motion and to non-Arrhenius 
behavior of thermal diffusion.
We have also pointed out the relation between deterministic and thermal 
diffusion. It would be interesting to enhance the 
complexity of the system by considering orientational degrees of freedom 
in two dimensions and anharmonic intramolecular potentials.

\section*{acknowledgements}
This work was supported by the Stichting Fundamenteel Onderzoek der 
Materie (FOM) with financial support from the Nederlandse Organisatie voor 
Wetenschappelijk Onderzoek (NWO), project 015.000.031. The authors wish 
to thank Ted Janssen for contributing to this work and Sergey Krylov for 
useful and stimulating discussions.

\end{multicols}


\begin{thebibliography}{99}
\bibitem{b.risken} H. Risken, {\it The Fokker-Planck Equation},
Springer, Berlin, {\bf 1989}, chapter 11.
\bibitem{b.kellogg} G. L. Kellogg, {\it Surf. Sci. Rep.} \textbf{1994}, 
{\it 21}, 1.
\bibitem{b.ala} T. Ala-Nissila, R. Ferrando, S. C. Ying, {\it Adv. Phys.}
{\bf 2002}, {\it 51}, 949.
\bibitem{b.venables} J. A. Venables, G. D. T. Spiller, M. Hanbucken, 
{\it Rep. Progr. Phys.} \textbf{1984}, {\it 47}, 399.
\bibitem{b.wang} R. Wang, K. A. Fichthorn, {\it Phys. Rev. B}
\textbf{1993}, {\it 48}, 18288.
\bibitem{b.braun} a) O. M. Braun, {\it Surf. Sci.} \textbf{1990}, 23, 262;
b) O. M. Braun, {\it Phys. Rev. E} \textbf{2001}, {\it 63}, 011102.
\bibitem{b.montalenti1} a) F. Montalenti, R. Ferrando, 
{\it Phys. Rev. Lett.} \textbf{1999}, {\it 82}, 1498;
b) F. Montalenti, R. Ferrando, {\it Phys. Rev. E} \textbf{2000}, {\it 61}, 
3411.
\bibitem{b.boisvert} G. Boisvert, L. J. Lewis, {\it Phys. Rev. B} 
\textbf{1997}, {\it 56}, 7643.
\bibitem{b.kovalev} A. S. Kovalev,  A. I. Landau, {\it Low Temp. Phys.}
\textbf{2002}, {\it 28}, 423.
\bibitem{b.romero} A. H. Romero, A. M. Lacasta and J. M. Sancho, 
{\it Phys. Rev. E} {\bf 2004}, {\it 69}, 051105. 
\bibitem{b.goncalves} S. Gon\c calves, V. M. Kenkre and A. R. Bishop, 
{\it Phys. Rev. B} {\bf 2004}, {\it 70}, 195415.
\bibitem{b.fusco1} C. Fusco, A. Fasolino, {\it Eur. Phys. J. B}
\textbf{2003}, {\it 31}, 95.
\bibitem{b.fusco2} C. Fusco, A. Fasolino, {\it This Solid Films}
\textbf{2003}, {\it 428}, 34.
\bibitem{b.deltour} P. Deltour, J.-L. Barrat, P. Jensen, 
{\it Phys. Rev. Lett.} \textbf{1997}, {\it 78}, 4597.
\bibitem{b.krylov} S. Yu. Krylov, {\it Phys. Rev. Lett.} \textbf{1999},
{\it 83}, 4602.
\bibitem{b.kurpick} a) U. K\"urpick, {\it Phys. Rev. B} \textbf{2001},
63, 045409; P. J. Feibelman, {\it Phys. Rev. B} \textbf{2000}, {\it 61}, 
R2452; b) F. Montalenti, R. Ferrando, {\it Surf. Sci.} \textbf{1999}, 
{\it 432}, 27.
\bibitem{b.hanggi} P. H\"anggi, P. Talkner, M. Borkovec, 
{\it Rev. Mod. Phys.} \textbf{1990}, {\it 62}, 251.
\bibitem{b.montalenti2} F. Montalenti, R. Ferrando, {\it Phys. Rev. B}
\textbf{1999}, {\it 59}, 5881.
\bibitem{b.hamilton} J. C. Hamilton, {\it Phys. Rev. Lett.} \textbf{1996},
{\it 77}, 885.
\bibitem{b.sholl1} D. S. Sholl, K. A. Fichthorn, {\it Phys. Rev. Lett.}
\textbf{1997}, {\it 79}, 3569.
\bibitem{b.sholl2} a) D. S. Sholl and C. K. Lee, {\it J. Chem. Phys.}
{\bf 2000}, {\it 112}, 817;
b) D. S. Sholl, {\it Ind. Eng. Chem. Res.} {\bf 2000}, {\it 39}, 
3737.
\bibitem{b.mitsui} T. Mitsui, M. K. Rose, E. Fomin, D. F. Ogletree,
M. Salmeron, {\it Science} \textbf{2002}, {\it 297}, 1850.
\bibitem{b.allen} M. P. Allen, D. J. Tildesley, {\it Computer Simulation of
Liquids}, Clarendon Press, New York, {\bf 1988}, pp. 185-187.
\bibitem{b.guantes} R. Guantes, J. L. Vega, S. Miret-Art\'es, 
{\it Phys. Rev. B} \textbf{2001}, {\it 64}, 245415.
\bibitem{b.sholl3} a) D. S. Sholl and R. T. Skodje, {\it Physica D} 
{\bf 1994}, {\it 71}, 168; b) D. Sholl and K. A. Fichthorn, {\it Phys. Rev. E}
{\bf 1997}, {\it 55}, 7753.
\bibitem{b.kopelevich} D. I. Kopelevich, H.-C. Chang, 
{\it Phys. Rev. Lett.} \textbf{1999}, {\it 83}, 1590.
\end{thebibliography}
\end{document}